# Atomically Thin Boron Nitride: Unique Properties and Applications

*Lu Hua Li* and Ying Chen**

Institute for Frontier Materials, Deakin University, Geelong Waurn Ponds Campus, VIC 3216, Australia

E-mail: luhua.li@deakin.edu.au, ian.chen@deakin.edu.au

Atomically thin boron nitride (BN) is an important two-dimensional (2D) nanomaterial, with many properties distinct from graphene. In this feature article, these unique properties and associated applications often not possible from graphene are outlined. The article starts with characterization and identification of atomically thin BN. It is followed by demonstrating their strong oxidation resistance at high temperatures and applications in protecting metals from oxidation and corrosion. As flat insulators, BN nanosheets are ideal dielectric substrates for surface enhanced Raman spectroscopy (SERS) and electronic devices based on 2D heterostructures. The light emission of BN nanosheets in the deep ultraviolet (DUV) and ultraviolet (UV) regions are also included for its scientific and technological importance. The last part is dedicated to synthesis, characterization, and optical properties of BN nanoribbons, a special form of nanosheets.

## 1. Introduction

Two-dimensional (2D) nanomaterials have attracted wide attention from both academics and industry since the discovery of graphene in 2004.[1] Graphene have many new and fascinating properties. The most famous case is that electrons act as mass-free Dirac particles in graphene. Although graphene is considered as one of the strongest materials, its structural anisotropy makes it highly flexible. There also exist many other 2D nanomaterials with diverse chemical compositions and complementary physical properties. Along with graphene, they are appealing building blocks for a new generation of electronic and optical devices. Similar to the case of graphene, atomic thickness can greatly change the physical and chemical properties of non-carbon 2D nanomaterials. Take $MoS_2$ as an example: with thickness reduction to monolayer, its bandgap not only increases but also changes from indirect to direct.[2,3]

Until now, most studies have focused on graphene, with little attention to atomically thin boron nitride (BN), or, in another name, BN nanosheets (here "nano" refers to thickness instead of lateral size). BN nanosheets have the same hexagonal structure as graphene but a white color, so they are sometimes called white graphene. The bond length of B−N is 1.44 Å,[4] and that of C−C in graphene is 1.42 Å. Thus, the lattice mismatch between graphene and BN is small (~1.6%). Due to their analogous crystal structures, atomically thin BN has many properties similar to those of graphene. Although the mechanical properties of monolayer (1L) BN have not been experimentally determined, the calculated Young's modulus of 1L BN is 0.71–0.97 TPa, and the breaking strength is 120–165 GPa.[5-8] These values are close to the experimental values of graphene (Young's modulus of 1.0 TPa and breaking strength of 130 GPa).[9,10] The thermal conductivity of few-layer BN was measured to be 100–270 W m$^{-1}$ K$^{-1}$,[11-13] making them one of the best electrically-insulating thermal conductors. Monolayer BN is expected to have larger thermal conductivity than few-layer BN,[14] but experimental verification is needed.

BN sheets also have many properties distinct from those of graphene, and it is important to explore and study these dissimilarities because they can lead to applications not possible from graphene. The most well-known difference between BN sheets and graphene is their electrical conductivity (and hence different color): graphene is a semi-metal, whereas BN sheets are insulating (according to theoretical calculations, the bandgap of BN sheets has a weak dependence on thickness). Therefore,

BN sheets can be used as dielectric substrates for graphene- and $MoS_2$-based heterostructures as electronic and optical devices.[15,16] However, many of these unique properties and applications of BN nanosheets have been much less recognized.

This feature article, based on the authors' recent research progress, presents an overview on the unique properties and applications of BN nanosheets that are not available to graphene. BN nanosheets have stronger resistance to oxidation than graphene, and hence are more suitable for applications at high temperatures and the manufacturing processes requiring heating treatments. The higher thermal stability and excellent impermeability of BN nanosheets make them superior to graphene in protecting metals from oxidation and corrosion. More importantly, the electrical insulation of BN nanosheets avoids galvanic corrosion that often happens between graphene and underlying metals. The thermal stability also makes BN nanosheets suitable as reusable substrates for surface enhanced Raman spectroscopy (SERS) with improved sensibility. Interestingly, the SERS signals from atomically-thin BN are stronger than those from bulk hexagonal BN (hBN) crystals due to better surface adsorption capability of atomically thin BN. BN nanosheets are also preferable dielectric substrates for graphene and other 2D nanomaterials; it is important to study the electrical field screening in BN nanosheets of different thicknesses for optimizing the performance of the 2D heterostructured devices. In contrast to graphene, which is not a luminescent material, BN nanosheets are efficient at emitting light, especially in the deep ultraviolet (DUV) and UV regions. The synthesis, structure, and optical properties of BN and graphene nanoribbons also merit comparison.

**2. Stronger Resistance to Oxidation**

Thermal stability is a fundamental property determining the temperature and environment under which a material can survive without dramatic physical and chemical failure due to oxidation and decomposition. Thermal stability is especially important to nanomaterials since a large portion of atoms in nanomaterials are sitting on the surface, and the oxidation kinetics become much higher. This is why the thermal stability of graphene in air has a strong dependence on thickness. Monolayer graphene starts to react with oxygen gas at 250 ºC, becomes heavily doped by oxygen at 300 ºC, and etched at 450 ºC; in contrast, bilayer and few-layer graphene nanosheets are not etched up to 500 ºC, and bulk graphite can withstand temperatures up to 800 ºC.[17] The oxidation of graphene not only affects its electrical property but also deteriorates its mechanical strength.[18] Thus, graphene is not suitable for high-temperature applications, or as a filler in metal or ceramic matrix composites whose fabrication normally requires high-temperature sintering treatment. In contrast, 1L BN can survive a much higher temperature (more than 800 ºC) in air.

To study their intrinsic thermal stability, BN nanosheets of different thicknesses were mechanically exfoliated from high-quality hBN single crystals on 90 nm oxide-covered silicon ($SiO_2$/Si) wafer. Optical microscopes were used to locate 1L and few-layer BN (**Figure 1**a).[19] It should be mentioned that the optical contrast for 1L BN on the $SiO_2$/Si wafer is only ~2.5% under white light, which is about one-quarter of that for graphene under the same condition.[20,21] Figure 1b shows the corresponding atomic force microscopy (AFM) image of the BN nanosheets. The AFM-measured heights of the exfoliated 1L BN is normally 0.4–0.5 nm, and 2L and 3L BN are 0.7–0.9 nm and 1.1–1.3 nm, respectively (Figure 1c).[22] These height values match those of graphene, as the interlayer spacings of BN and graphite are close.

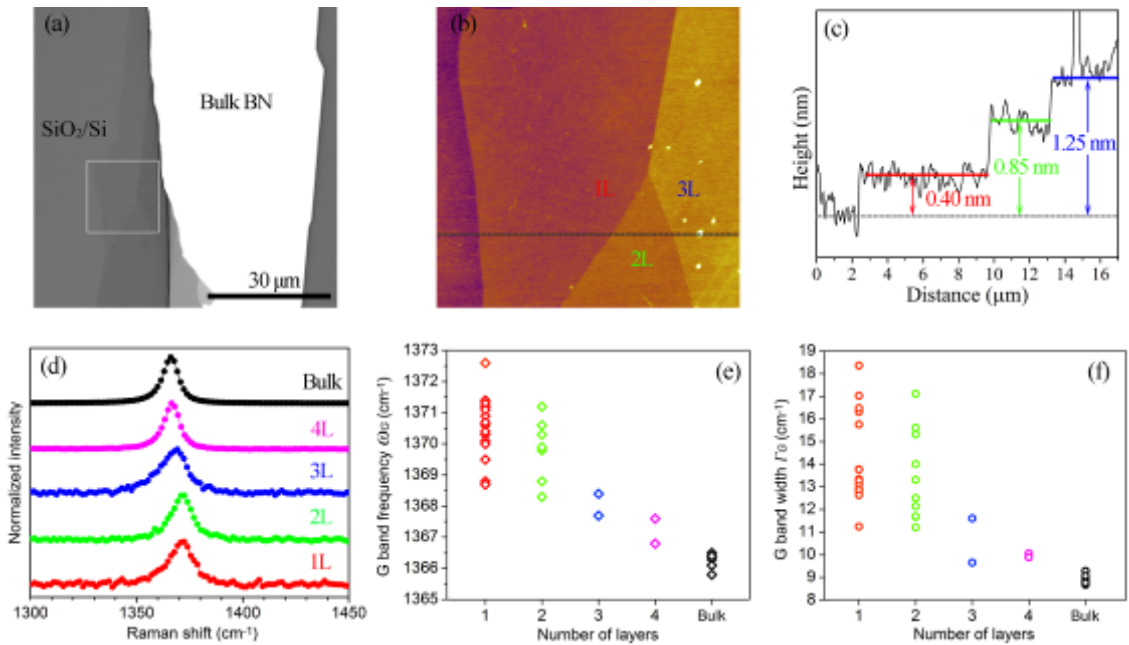

**Figure 1**. (a) Optical microscopy image of 1–3L BN on SiO$_2$/Si substrate; (b) the corresponding AFM image; (c) AFM height traces of the 1–3L BN; (d) typical Raman spectra of high-quality BN nanosheets and a bulk hBN single crystal; (e), (f) summaries of the Raman G band frequency and width of BN nanosheets of different thicknesses.[19,22]

Raman spectroscopy is a valuable tool to determine the thickness and crystallinity of graphene due to the characteristic G, D, and 2D bands, but BN only exhibits a Raman G band corresponding to the E$_{2g}$ vibration mode. No D band is detectable from BN materials because of the lack of Kohn anomaly. In addition, the Raman signals from BN are much weaker than those from carbon materials. Nevertheless, Raman spectroscopy still provides useful information on the thickness and quality of BN nanosheets. Figure 1d compares the normalized G bands of the exfoliated BN of different thicknesses. The frequency of the G band upshifts with decreased thickness of BN nanosheets, which can be better seen in Figure 1e. The G band frequency for 1L BN ($N$=15) sits between 1368.6 cm$^{-1}$ and 1372.7 cm$^{-1}$ with an average value of 1370.5±0.8 cm$^{-1}$; 2L BN is in the range of 1368.2 and 1371.3 cm$^{-1}$ with an average of 1370.0±0.6 cm$^{-1}$ ($N$=6); those of 3L and 4L BN nanosheets are close to 1368 cm$^{-1}$ and 1367 cm$^{-1}$, respectively. The G band frequency for bulk crystals is 1366.2±0.2 cm$^-$ 1.[22] As BN has no doping effect from the substrate, the Raman frequency change should be due to higher strain in BN nanosheets caused by their much lower rigidities,[23] the uneven $SiO_2$ substrate,[24] and lower interlayer interaction;[25] all of these factors result in phonon hardening and G band upshifts. The full widths at half-maximum (FWHM) of the Raman G band increase with reduced BN thickness (Figure 1f). Similarly, this can be attributed to the increased strain and stronger surface scattering in BN nanosheets, as these factors decrease vibrational excitation lifetime and hence increase G bandwidth. It should be mentioned that Gorbachev *et al.* reported G band upshifts for 1L and downshifts for 2–6L BN,[21] and it is still unclear what caused the discrepancy.

To investigate the intrinsic thermal stability of BN nanosheets of different thicknesses, the materials were heated at different temperatures for 2 h on $SiO_2$/Si wafer at different temperatures in an open-air environment. After heating to 840 ºC, BN nanosheets, including the monolayer, did not show any morphology change or etching according to the AFM measurements of 1L BN (**Figure 2**a). Etching appeared on 1L BN at 850 ºC (Figure 2b). At the same temperature, 2L and 3L BN nanosheets were visually intact (Figure 2c and e). After heating at 860 ºC, 1L BN was burnt out, and etch pitches were observed on 2L and 3L BN (Figure 2d and f). BN of 2L and 3L were burnt out at 870 ºC. Lateral sizes (all in micron range) seemed not to affect the oxidation behavior. These results indicate that, similar to graphene,[17,26] the chemical reactivity of BN depends on thickness, but the oxidation resistance of BN nanosheets exceeds that of graphene, and the oxidation temperature among BN nanosheets of different thicknesses is much less thickness-dependent (only 10 ºC difference between 1L and 2L BN, in comparison to 50 ºC difference between 1L and 2L graphene).[17] Thus, BN nanosheets can be used at higher temperatures.

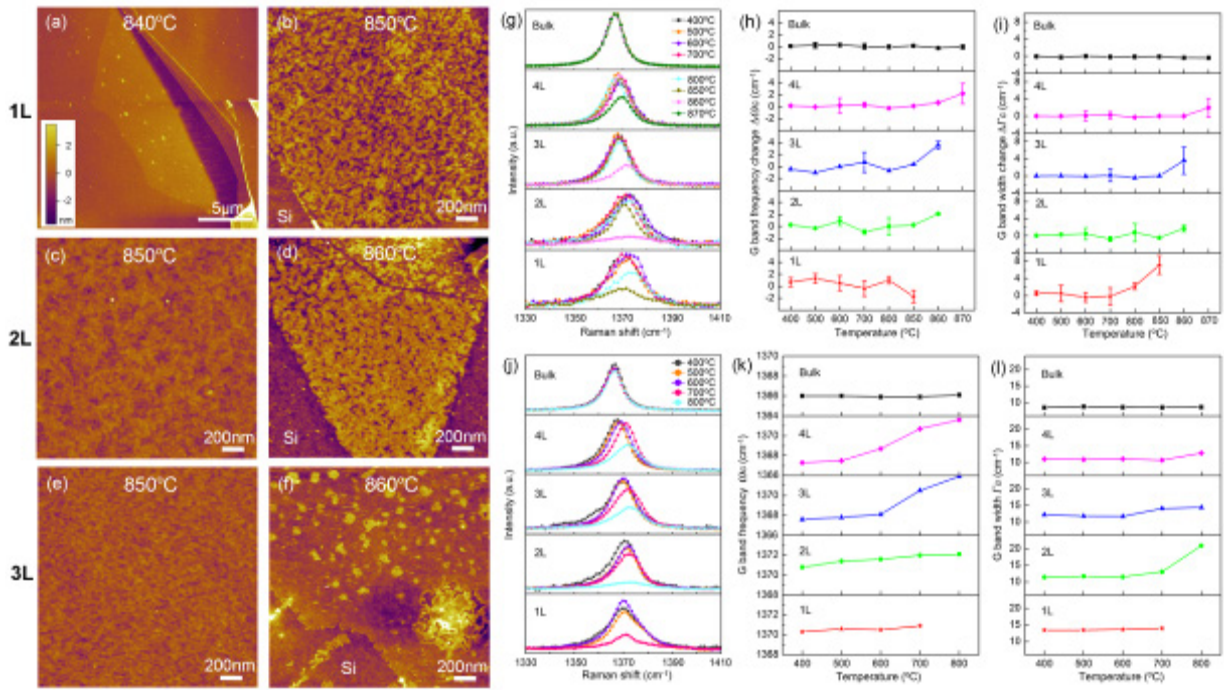

**Figure 2.** (a–f) AFM images of 1–3L BN on SiO$_2$/Si after heating at different temperatures in air for 2 h; (g–i) Raman spectra of BN nanosheets of different thicknesses after heating treatments at different temperatures, and the corresponding changes of the G band frequency and width; (j–l) Raman spectra of 1–4L BN after sequential heating treatments and the corresponding changes of the G band frequency and width.[22]

Oxygen doping in the BN nanosheets after heating treatments at different temperatures was examined using Raman spectroscopy. Oxygen doping can occur to BN nanosheets without noticeable morphology change but affect their properties. For example, when BN nanosheets are used as substrates for graphene or MoS$_2$ nanosheets, it creates pinholes and causes current leakage and device failure. Figure 2g compares the Raman spectra of monolayer, few-layer, and bulk BN after heating at various temperatures between 400 °C and 870 °C for 2 h in air. For the 1L BN, the intensity of the G band decreased noticeably from 800 °C, suggesting strong oxidation from this temperature. Dramatic decrease of the Raman G band occurred to bilayers and trilayers after 860 °C heating treatment. The oxidation also influenced the frequency and FWHM of the Raman G band (Figure 2h and i). These results show that 1L BN had no oxygen doping at less than 700 °C in air, and 2L and 3L BN are not doped until 800 ºC. The stronger oxidation resistance of BN nanosheets can be ascribed to the higher oxygen adsorption energy than desorption energy, as reflected by the high oxygen doping temperature shown by the Raman results.

Repeated or extended heating treatments can reduce the thermal stability of BN nanosheets. This was tested by heating one nanosheet sample at 400 ºC in air for 2 h, then 500 ºC for 2 h, up to 800 ºC. Monolayer BN burnt out after the sequential heating up to 800 ºC (total heating time 10 h). The changes of the Raman G band frequency and FWHM of 1–4L and bulk BN are shown in Figure 2j, k, and l. Following the same analyzing principle, it can be concluded that oxygen doping happened at a lower temperature for few-layer BN, i.e. above 600 ºC. Thus, similar to graphene, repeated and extended heating in air reduces the thermal stability of BN nanosheets, probably due to reduced activation energy for oxidation caused by increased roughness and more bonding distortions in BN nanosheets.

Oxidized graphene showed hexagonal or close-to-round etch pits, implying radial etching from defects. However, the etch pits in the oxidized BN nanosheets were either elongated or randomly shaped. This difference can be attributed to the high crystallinity of the BN nanosheets, which have few defects. The elongated etch lines agree with the oxidation mechanism predicted by theoretical calculations: oxygen chains are formed on 1L BN, and then the dissociation of N−O bonds cuts the monolayer in a straight line.[27,28]

BN nanosheets of lower quality show slightly lower oxidation temperatures. **Figure 3**a and b display the AFM height and phase images of a 1L BN nanosheet exfoliated from commercial hBN particles (PT110, Momentive), which contain more defects than hBN single crystals, and the sample was heated at 750 ºC in open air for 2 h. Because of the atomic thickness of 1L BN, the AFM phase image is better than height image at showing possible etching traces after heating treatment. There was no noticeable etching of the nanosheet after heating at 750 ºC. A 2L BN was heated at 775 ºC for 2 h, and etch pits appeared after the heating treatment, as shown in Figure 3c and d. In this case, the oxidation initiated at the defects, visible as round shapes for most of the small etch pits; however, larger etch pits became more elongated, suggesting that the oxidation mechanisms in BN and graphene are different. Therefore, the thermal stability of BN nanosheets depends on their crystal quality, i.e. content of defects.

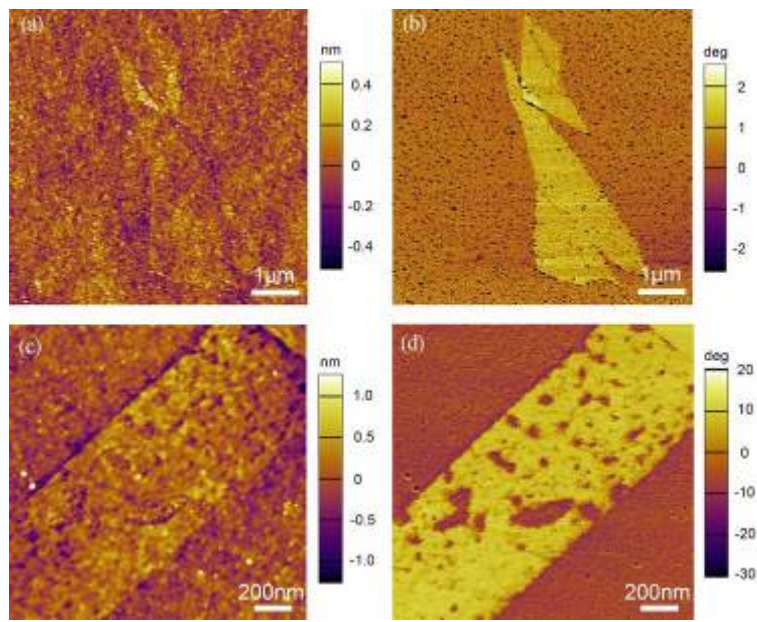

**Figure 3.** AFM height and phase images of (a, b) a 1L BN of lower quality after heating at 750 °C in air for 2 h, and (c, d) 2L BN after heating at 775 °C under similar conditions.

## 3. Better Choice as Protective Barrier

Graphene is highly impermeable to gas and moisture, and non-reactive to most chemicals, so it has been proposed as a surface coating on metals to prevent oxidation and corrosion. Although excellent protection could be provided by graphene in the short term,[29,30] graphene enhances the corrosion of the underlying metal in the long term because it acts as a galvanic cell.[31,32] The excellent impermeability, superb mechanical strength, and good thermal conductivity make BN nanosheets a potential barrier to protect metals from oxidation and corrosion as well. Most importantly, BN nanosheets are electrical insulators and thus do not cause galvanic corrosion.[33] In addition, BN nanosheets have higher thermal stability than graphene, and hence can function at much higher temperatures.

Copper foils covered by 1L and ~20L BN nanosheets grown by chemical vapor deposition (CVD) were used to examine their protective performance against thermal oxidation in air and corrosion in NaCl solution, and a bare Cu foil was used as a control sample.[33] The oxidation resistance of the Cu foils protected by the BN nanosheets was tested by heating the samples at 250 ºC in open air for up to 100 h. Before heating, the BN-covered Cu foils had a metallic color similar to that of the bare foil, due to the atomic thickness and low absorption of visible light of BN nanosheets (insets of **Figure 4**a, d and g). The bare Cu and 1L BN-covered Cu foils became dark brown after heating for 2 h (insets of Figure 4b and e), and the corresponding scanning electron microscope (SEM) images show large numbers of particles (cupric oxide and cuprous oxide), indicating severe oxidation of their surfaces. In contrast, the metallic luster was retained on the 20L BN foil after the same heating treatment, although the optical image reveals black spots (inset of Figure 4h). Thus, the majority of the surface of 20L BN-covered foil was intact (Figure 4i). After heating for 20 h and 100 h, the bare and 1L BN-covered Cu foils became almost black (insets of Figure 4c and f), and the oxide particles grew larger (Figure 4c and f), whereas most of the surface of the 20L BN foil became only slightly darker (inset of Figure 4i), and decorated with some tiny oxide particles (Figure 4i). The different degrees of oxidation of the three samples were confirmed using energy-dispersive X-ray (EDX) spectroscopy (Figure 4j). For the bare Cu foil, the atomic ratio between O and Cu increased dramatically from 0.021±0.005 to 0.554±0.095 just after 2 h of heating treatment. The 1L BN-covered foil showed a similar trend. In contrast, the O:Cu ratio of the 20L BN foil only showed a modest increase, from 0.024±0.004 to 0.088±0.020, even after heating for 100 h. In other words, the 20L BN reduced ~90% of the oxidation of Cu foil. Therefore, the 1L BN had no protection, whereas the thicker CVD BN nanosheets did show a good but incomplete protection.

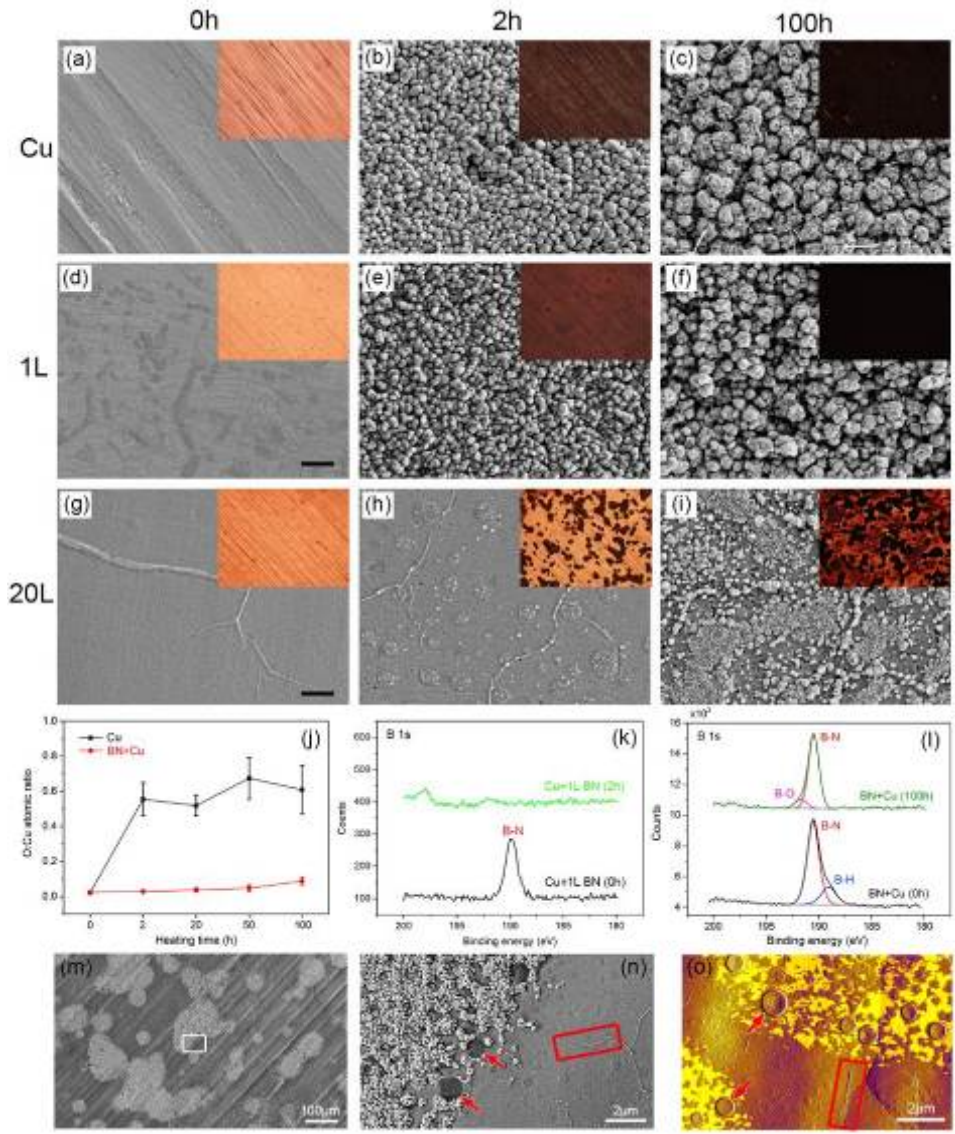

**Figure 4.** (a–i) SEM images of bare Cu foil, and Cu foil covered by 1L and 20L CVD-grown BN nanosheets after heating at 250 °C for different time lengths, with corresponding optical images inserted; (j) Oxygen contents, calculated by EDX, of three samples after oxidation; (k, l) XPS spectra of 1L and 20L BN after oxidation for 2 h and 100 h, respectively; (m, n) SEM images of the uneven protection of 20L BN after 2 h oxidation; (o) overlay of AFM deflection image and conductive AFM current mapping of the 20L BN-covered Cu foil without oxidation. Copyright © 2014 WILEY-VCH Verlag GmbH & Co. KGaA, Weinheim.[33]

The incomplete protection of the 20L BN was found to be due to its inhomogeneous quality and crystallinity.[33] The CVD-grown 20L BN showed either weak and broad G band at 1370 cm$^{-1}$ or no G band at all, indicating its low structure quality. Conductive AFM was used to map the quality of the nanosheet. High-quality BN nanosheets are electrically insulating; however, impurities and defects increase their conductivity. Figure 4o shows the overlay of surface morphology (deflection image) and current mapping (conductive AFM) of the 20L BN-covered Cu foil. The deflection image shows some BN disks (arrows) and ripples of BN nanosheet (rectangles). The conductive AFM image shows no current at the areas close to the ripples and currents greater than 8 nA at the areas around the BN disks (yellow), so the quality of the BN nanosheet around the BN disks was relatively worse than at the rippled regions. The SEM image of the 20L BN-covered foil after heating for 2 h shows that the oxide particles formed in the BN disk areas but not in the rippled regions (Figure 4m and n). It is clear that the black spots on the 20L BN foil after the heating treatment (insets of Figure 4h and i) were due to the low quality of these areas, where oxygen can penetrate through the BN nanosheet and oxidize the underlying Cu foil.

The unsatisfactory protection of the CVD-grown 1L BN is also due to its low quality. X-ray photoelectron spectroscopy (XPS) analyses revealed that the 1L BN foil before heating shows a peak at ~190 eV in the B 1s region (black in Figure 4k), confirming the presence of the B–N bond. However, after heating for 2 h, the B–N peak was absent (green in Figure 4k), suggesting destruction of the 1L BN. Raman spectroscopy confirmed the bad quality of the CVD 1L BN. However, the disappearance of the 1L CVD BN nanosheet was not due to its oxidative decomposition. First, although the oxidation temperature of BN nanosheets depends on quality/crystallinity, 1L BN should not decompose at 250 ºC in air. Second, when the 1L BN was transferred to SiO$_2$/Si substrate using a polymer-assisted technique and heated under the same condition, it did not show any morphology change or sign of oxidation. Therefore, the loss of the 1L BN after 2 h heating treatment in air appears to be due to diffusion of oxygen through the low-quality nanosheet, and then the formation of oxide particles, which tear up the atomically-thin BN coating. After being broken into tiny pieces, the 1L BN nanosheet became much more easily oxidized and it eventually burned out. This process was similar to that proposed for the destruction of graphene on Cu foil after a long-term oxidation at room temperature.[32] In contrast, the B–N XPS peak from the 20L BN only had a little reduction in intensity, in spite of the appearance of a shoulder peak of the B−O bond at 191.7 eV after 100 h heating (Figure 4l).

BN nanosheets can also be used to protect other 2D nanomaterials. Graphene and $MoS_2$ transistors sandwiched by BN nanosheets can be used at higher temperatures. Such protection is more critical to air-sensitive 2D nanomaterials such as phosphorene, which degrades in an ambient environment. Phosphorene encapsulated by BN nanosheets becomes not only stable under atmospheric conditions, but also free of surface charges so that much better mobility can be achieved.[34] However, as shown above, the protection performance of BN nanosheets highly depends on crystal quality. Therefore, large-scale growth of high-quality BN nanosheets is needed. Such a synthesis technique is also important to graphene electronics and many other applications.

The oxidation protection of metal substrate also provides an easy and straightforward way to test the quality of CVD-grown BN nanosheets. Usually, CVD-grown BN nanosheets have to be transferred to $SiO_2$ substrates following a complicated and time-consuming process so that their quality can be determined by mechanical or electrical measurements.[35,36] According to the results of the present study, the quality of CVD-grown BN nanosheets can be easily evaluated by heating treatment in air. For example, the area of the black regions on the 20L BN foil was about 30%, which quantifies the percentage of the bad-quality area in the nanosheet. In addition, the higher the temperature the metal foil can sustain, the better the quality of the CVD-grown BN is.

The anti-corrosion performance of the CVD-grown 1L and 20L BN nanosheets was further investigated electrochemically in NaCl solution (0.1 M).[33] Open circuit potentials of the three different samples are compared in **Figure 5**a. The 20L BN foil had the highest corrosion potential in the NaCl solution and reached equilibrium most quickly, and therefore is the most stable among the three samples. In comparison, the 1L BN foil showed a small drop of potential first, and then a gradual decrease to the equilibrium potential at about −160 mV (blue in Figure 5a). The potential of the bare Cu foil first decreased, and then stabilized at about −165 mV (black in Figure 5). The smaller decreases of the open circuit potentials suggest that both BN nanosheets protected the underlying Cu foils from oxidation in the electrochemical environment.

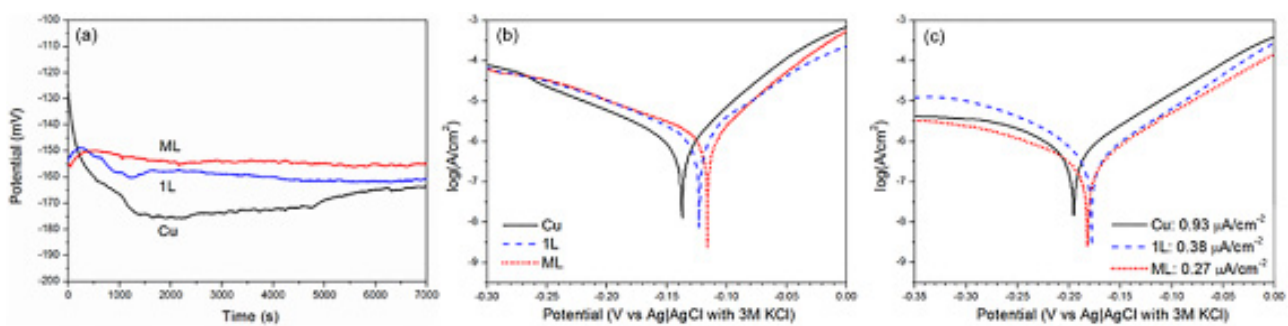

**Figure 5.** (a) Open circuit potentials of bare Cu foil and Cu foil covered by 1L and 20L(ML) CVD-grown BN nanosheets in NaCl solution (0.1 M); Tafel plots in aerated (b) and nitrogen gas-saturated (c) NaCl solution. Copyright © 2014 WILEY-VCH Verlag GmbH & Co. KGaA, Weinheim.[33]

Tafel results are shown in Figure 5b and c. Lower anodic currents but higher cathodic currents were observed from both BN nanosheet-covered Cu foils in aerated NaCl solution (Figure 5b). It implies that the BN nanosheets protected the Cu foils from oxidation but enhanced oxygen reduction reactions. Similar cyclic voltammetry tests were also conducted in nitrogen-saturated NaCl solution (Figure 5c). It confirmed that the 20L BN nanosheet did protect the underlying Cu, evidenced by the lower anodic and cathodic currents, decreased corrosion current density, and corrosion potential shift. Based on these results, even though the oxygen reduction can accelerate the oxidation of Cu, the lower anodic current from the 20L BN foil in the aerated solution suggests that the oxygen reduction by the BN nanosheet did not enhance the formation of cuprous oxide on the underlying Cu surface. It is expected that better protection can be achieved when BN nanosheets of higher quality are used, as the oxygen reduction reactions may come from the impurities and defects of the CVD-grown BN nanosheets.

## 4. Reusable Substrates for Surface-Enhanced Raman Spectroscopy

The high thermal stability and excellent adsorption capability of BN nanosheets enable highly-efficient and reusable sensors. Efficient substrates of BN nanosheets decorated by noble metal nanoparticles can be used for surface enhanced Raman spectroscopy (SERS).[19] The plasmonically-active Au particles were produced by a straightforward and controllable method: sputtering and thermal annealing. The particle size could be well controlled by adjusting sputtering thickness and annealing temperature, and the reproducibility is high.

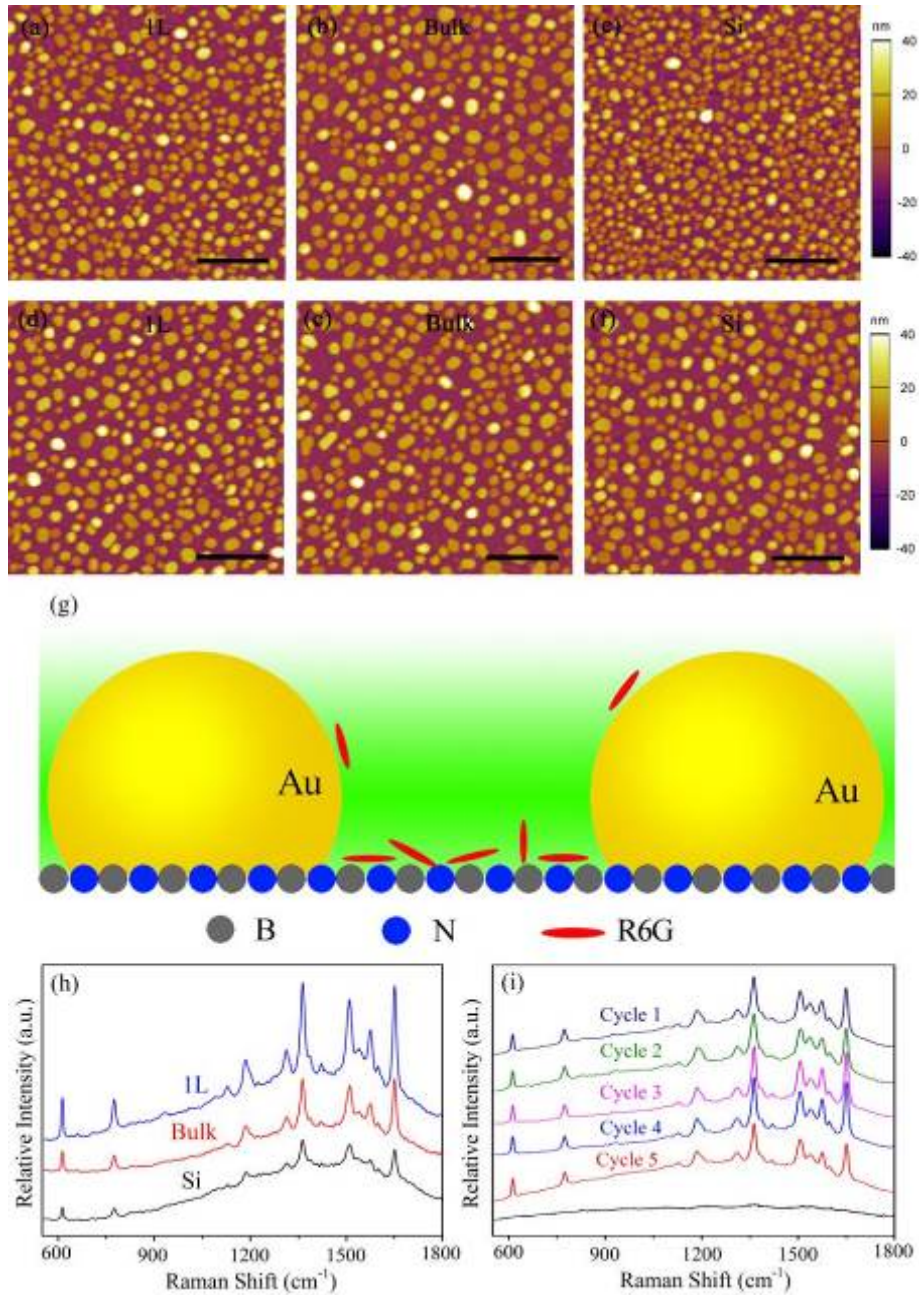

**Figure 6.** AFM images of Au particles produced by the sputtering and annealing method on surfaces of (a) 1L BN, (b) bulk hBN, and (c) SiO$_2$/Si; (d–f) AFM images of Au nanoparticles of similar sizes and distributions on the three surfaces; (g) illustration of effects of BN nanosheets on enhancing Raman signals; (h) comparison of SERS spectra of R6G ($10^{-6}$ M) on 1L BN, bulk hBN, and SiO$_2$/Si; (i) reusability tests of the SERS substrate using 1L BN.[19]

It was observed that the size of the Au particles that formed on the surfaces of atomically thin BN, bulk hBN crystal, and SiO$_2$/Si wafer were different, even under the same sputtering and annealing condition. The particles on BN were slightly larger than those on the SiO$_2$/Si wafer (**Figure** 6a-c) due to a larger surface diffusion coefficient on BN, which means higher migration rates of Au atoms. Nevertheless, the particle size on the 1L BN was much smaller than that on the bulk hBN. A similar phenomenon has been reported for graphene.[37] This was due to the larger roughness of atomically thin BN caused by the uneven SiO$_2$ substrate, which reduced the diffusion coefficient and hence the particle size.

To compare the Raman enhancements of atomically thin BN, bulk hBN, and SiO$_2$/Si substrates, Au particles of a comparable size and density were produced on the three substrates using slightly different sputtering times (Figure 6d–f). The samples were immersed in rhodamine 6G (R6G) solution ($10^{-6}$ M) for Raman spectroscopy testing. Because of the better affinity of aromatic molecules to BN than SiO$_2$ (Figure 6g), both atomically thin and bulk BN had better Raman enhancements than the SiO$_2$/Si substrate (Figure 6h). Intriguingly, the Raman signals of R6G were slightly stronger from the 1L BN than from the bulk hBN (Figure 6h). The different signal strength cannot be attributed to a plasmonic effect, as the size and density of Au particles on the two substrates were tuned to the same level (Figure 6d–f). Neither could it be due to different chemical enhancements of BN of different thickness.[38] Instead, it implies different adsorption capabilities of BN of different thicknesses. According to the authors' recent studies, nanosheets do have better adsorption per unit surface area than the corresponding bulk crystals due to conformational change. It should also be mentioned that the Au particles seem not to enhance the Raman signal of BN, and the Raman G band of atomically thin BN is so weak that it is not noticeable and hence does not interfere with the signals of the analyte. In contrast, if graphene is used as a SERS substrate, its Raman bands are normally present.

BN nanosheets can withstand higher temperatures than most aromatic molecules, so the BN nanosheet-based SERS substrates can be reused by removing the adsorbed molecules by oxidation in air. In contrast, graphene does not have such reusability. The reusability of a high-quality 1L BN with Au particles is shown in Figure 6i. In each cycle, the adsorbed R6G molecules were cleaned by heating at 400 ºC in air for 5 min. After five cycles of heating and re-adsorption, there was no loss of SERS enhancement.

We expect improved enhancement factor if BN nanosheets are used to cover metal particles because the adsorption area dramatically increases. Such design also has other benefits. Silver particles are more effective than Au in Raman enhancement of many molecules such as R6G with excitation of 400-600 nm,[39] but are prone to oxidation and hence loss of enhancement. Although graphene has been proposed to veil Ag particles for protection,[40,41] but graphene enhances the oxidation of the Ag nanoparticles due to galvanic corrosion, similar to the Cu case described previously. Therefore, it is advantageous to use BN nanosheets instead of graphene for oxidation protection. In addition, BN can sustain high temperatures in air, so Ag nanoparticles covered by BN nanosheets can be cleaned by heating in air to achieve reusability. By using large-sized CVD-grown BN nanosheets, such BN-veiled SERS substrates can be easily scaled up to wafer sizes using the sputtering and annealing process. Again, the reusability and stability highly depends on the quality of nanosheets.

## 5. Excellent Dielectric Substrate and Electric Field Screening

BN nanosheets are excellent substrates for graphene, $MoS_2$ nanosheets and other 2D nanomaterials in electronic and optical applications[16,42-44] because of their large bandgap, surface flatness, and freedom from surface charges. In addition, the atomic thickness enables BN nanosheets to form heterostructures with other 2D nanomaterials. Electric field screening in BN nanosheets of different thicknesses could affect transport properties and gate voltage control of graphene/BN structures, and it also changes the dielectric property of graphene above.[45] Traditionally, the screening property of a material is determined by capacitor and optical methods. However, it is difficult to use these methods on BN nanosheets due to their atomic thickness. To overcome this challenge, electric force microscopy (EFM) was used to study the dielectric screening in BN nanosheets. EFM, a technique derived from scanning probe microscopy, is sensitive to surface charges. When BN nanosheets are placed on $SiO_2$, dipolar water films form between the interfaces of the two materials, and can act as an external electric field.[46] EFM can be used to detect the electric fields through BN nanosheets of different thicknesses and thus to deduce different dielectric screening properties.[47]

BN nanosheets were mechanically exfoliated from hBN single crystals onto $SiO_2$/Si substrates. **Figure 7**a shows an AFM image of the sample, which contains 1–24L BN nanosheets. In the measurements, EFM phase images were recorded when different cantilever tip voltages were applied. Under a tip voltage of +6 V, BN nanosheets showed more positive EFM phase shifts than the $SiO_2$ substrate (Figure 7b), whereas more negative phase shifts were recorded from the BN nanosheets under negative tip voltages (Figure 7c). These results confirm that there existed a dipolar water film between the BN nanosheets and $SiO_2$ substrate. In addition, BN nanosheets of different thicknesses showed different EFM phase contrasts under different tip voltages. Monolayer BN showed more EFM phase shifts than thicker BN nanosheets under +6 V tip voltage (Figure 7b), but the EFM phase contrast reverted under –6 V (Figure 7c). According to the theory of EFM, the different phase shifts represent different unscreened surface charges (i.e. different unscreened electric fields) on BN nanosheets of different thicknesses.

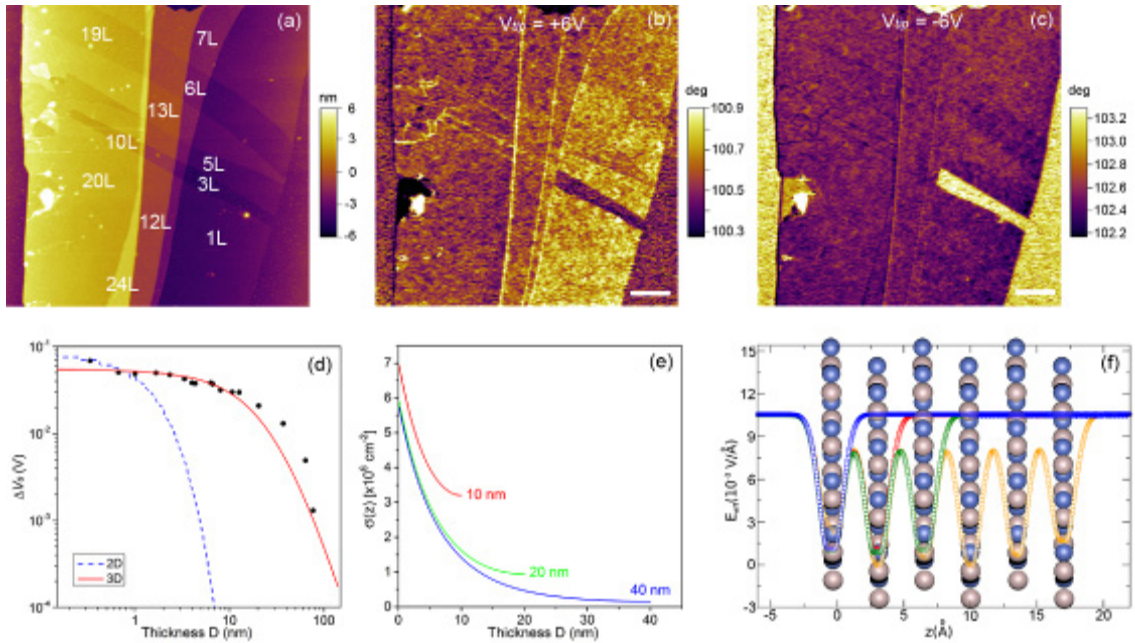

**Figure 7.** (a) AFM image of 1–24L BN nanosheets on SiO$_2$/Si; corresponding EFM phase images under tip voltages of (b) +6 V and (c) −6 V; (d) EFM-deduced different electric field screening properties of BN nanosheets of different thicknesses, along with the fittings using non-linear Thomas–Fermi theories (blue: 2D model; red: 3D model); (e) screening inside BN nanosheets of 10 nm, 20 nm, and 40 nm thickness, calculated from the fittings with interlayer hopping under consideration (i.e. 3D model); (f) DFT-calculated electric field distributions in 1L, 2L, 3L, and 6L BN nanosheets under an external electric field. Adapted with permission from *Nano Letters*. Copyright (2014) American Chemical Society.[47]

The screening properties of BN nanosheets can be quantified by plotting out the EFM phase values versus tip voltages. The comparison of the EFM phases of 1L, 2L, and 24L BN nanosheets did show shifts, indicating their different screening properties. The EFM results are summarized in Figure 7d. It can be concluded that the dielectric screening in BN decreases with its thickness decrease. However, such a decrease in screening in BN is much less steep than those in graphene and MoS$_2$.[48,49] In other words, the electric field screening in BN is much less thickness-dependent.

Non-linear Thomas–Fermi theory was used to gain insight to the screening properties of BN. The blue dotted line in Figure 7d was the fitting curve using this theory without consideration of the interlayer hopping effect (i.e. 2D model). Such fitting shows that the dielectric screening in BN should drop sharply with the decrease of thickness. However, it does not match the experimental data. Therefore, interlayer hopping between BN layers has to be considered. The red line in Figure 7d shows that fitting using non-linear Thomas–Fermi theory with the hopping effect included (i.e. 3D model) is in much better agreement with the EFM results: BN is mostly in a strong coupling regime where the electrostatic energy surpasses the kinetic energy. The charge density distributions in BN nanosheets of different thicknesses were also calculated based on the fitting results from the theory, and are shown in Figure 7e. However, non-linear Thomas–Fermi theory has a continuum limit so that it is only valid when the thickness of BN nanosheet is much larger than its interlayer distance (0.334 nm). Thus, this theory is not suitable for atomically thin BN nanosheets. Therefore, first-principles total-energy calculations based on density functional theory (DFT) were performed to better understand the dielectric screening in atomically thin BN. In the calculations, 1–6L BN nanosheets were placed between an external electric field (0.0105 V/Å), and the effective electric field between their layers was deduced, as shown in Figure 7f. The effective electric field in BN of different thicknesses showed a similar distribution: the maximum was located at the center of two BN layers, and for 1–6L BN nanosheets, the values were comparable. It is consistent with the EFM results that the dielectric screening in BN has a weak dependence on thickness. The calculated dielectric constants for 1L, 2L, and 3L BN nanosheets were 2.31, 2.43, and 2.49, respectively. This study not only reveals a fundamental property of BN nanosheets, but also guides the design and optimization of 2D heterostructured electronic and optoelectronic devices using BN nanosheets.

## 6. Special Luminescence

Because of the lack of bandgap, intrinsic graphene has no luminescence. In contrast, BN crystals and nanostructures are efficient in light emission, especially in the DUV region. DUV light has a wide range of applications, including in medical science, electronic device, data storage and nano-fabrication. Although the luminescence of bulk hBN crystals[50-55] and BN nanotubes[56-59] has been systematically studied, there are only a few experimental studies on the luminescence of BN nanosheets.

Synchrotron-based photoluminescence spectroscopy was used to study the light emission of BN nanosheets.[60] Because the measurements required a large quantity of BN nanosheets, a tailored ball milling method was used to produce adequate samples. Commercial hBN powder (diameters 0.3–1 µm and thicknesses 20–110 nm) was ball milled under gentle shear force in benzyl benzoate.[4] The shear force can exfoliate bulk crystals to nanosheets with little damage to in-plane structure due to much weaker interlayer interactions, as shown by the two schemes in **Figure 8**a. Benzyl benzoate was used for several reasons. First, it can greatly reduce the mechanical impact on hBN crystals from the milling balls; second, it prevents the re-stacking of the newly-formed BN nanosheets due to the similar surface tension of benzyl benzoate to that of BN;[61] third, it dramatically reduces the milling contamination. Other solvents can also be used if these requirements are met.

Figure 8b shows a typical SEM image of the BN nanosheets produced after ball milling for 15 h and centrifugation for removal of thick particles. The thickness of the BN crystals was remarkably reduced during the process, as the produced nanosheets were almost transparent to the electron beam. The morphology and crystallinity of the BN nanosheets were examined by transmission electron microscopy (TEM). Figure 8c shows a TEM image of a BN nanosheet, and the corresponding electron diffraction pattern suggests highly crystalline structures, with a certain amount of stacking faults (inset of Figure 8c). The excellent crystallinity was confirmed from a high-magnification TEM image, which shows hexagonal lattice dots typical for hBN materials (Figure 8d). Figure 8e and f show the edges of 3L, 4L, and 5L BN nanosheets. Bilayer BN was also observed during TEM investigations.

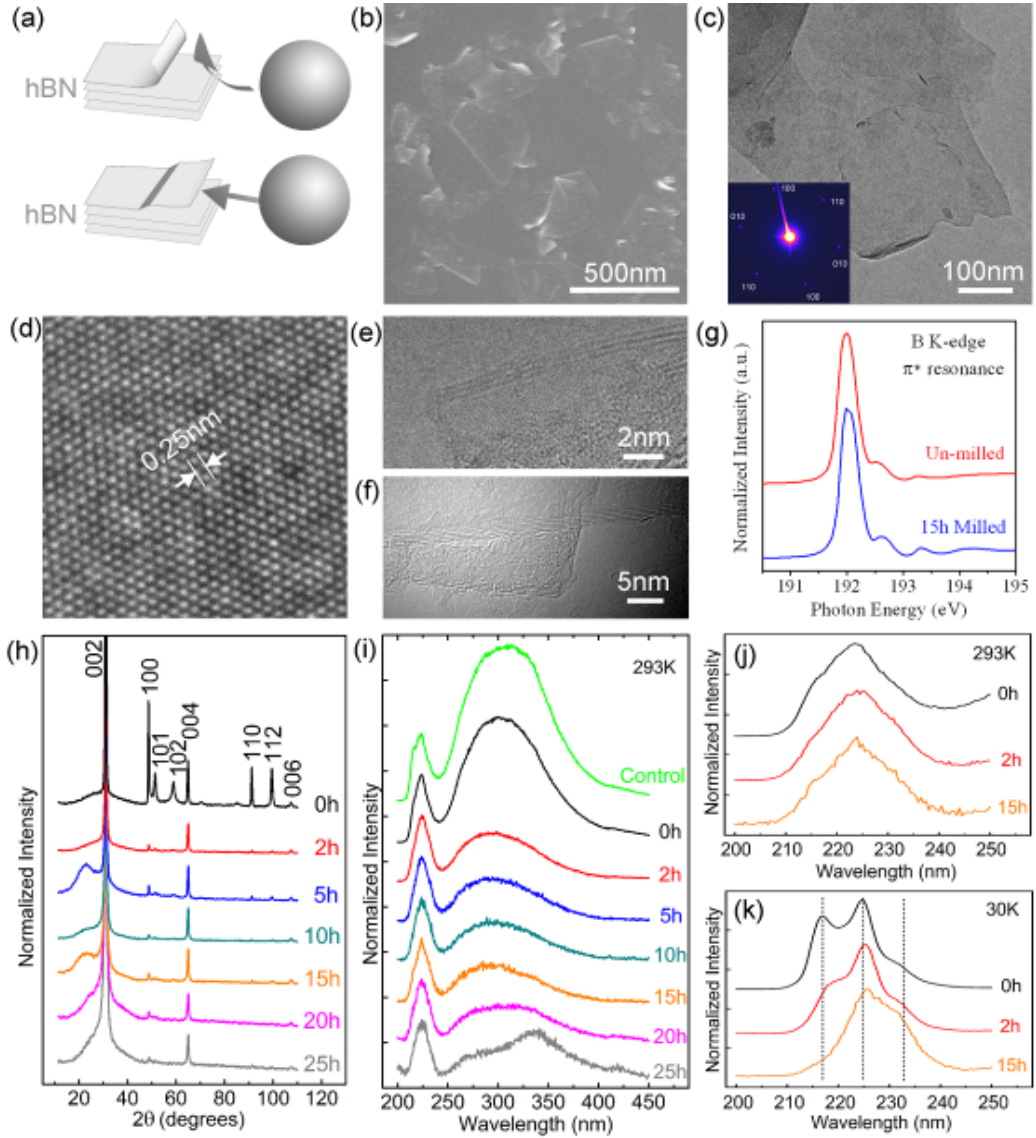

**Figure 8.** (a) Mechanism of nanosheet production by ball milling method; (b) SEM image of synthesized BN nanosheets; (c) TEM image of a BN nanosheet, with the corresponding diffraction pattern inserted; (d) high-magnification TEM image with the lattice fringes of BN shown; (e, f) TEM images of edges of three BN nanosheets; (g) comparison of NEXAFS spectra of the starting hBN powder and BN nanosheets produced by ball milling method; (h) XRD and (i) PL spectra of BN nanosheets after ball milling treatment for different time lengths; (j, k) room-temperature and 30 K PL spectra in the deep UV region of the starting hBN, and BN nanosheets ball milled for 2 and 15 h.

Defects and chemical impurities, which can dramatically affect the optical properties of BN nanosheets, were probed by near-edge X-ray absorption fine structure (NEXAFS) spectroscopy. NEXAFS spectroscopy is especially useful to study BN materials because of well separated $\pi^*$ and $\sigma^*$ resonances in B K-edge region.[62-64] The NEXAFS spectra of the starting hBN particles and the BN nanosheets produced by ball milling are compared in Figure 8g. The spectrum of the starting particles mainly shows a strong and sharp $\pi^*$ resonance at 192.0 eV, corresponding to the B 1s-$\pi^*$ transitions that meet the dipole selection rule. The three satellite resonances at 192.6, 193.3, and 194.1 eV are due to pre-existing defects, i.e. B–2N–O, B–N–2O, and B–3O, respectively.[62-64] The NEXAFS spectrum of the BN nanosheets is similar to that of the starting material, with only a slight increase in the three satellite peaks, which indicates that the BN nanosheets have a density of defects and impurities comparable to the starting commercial hBN powder. In other words, the ball milling process reduces the thickness of BN particles without introducing many defects to the crystal structure. The effect of the gentle milling treatment is also reflected in the X-ray diffraction (XRD) results in Figure 8h.[60] The XRD pattern of the starting material shows typical diffraction peaks of hBN structure. After ball milling for just 2 h, only the (002) and (004) peaks remain in the pattern, and the relative intensity of the (004) peak is slightly decreased. With the increase of the ball milling time to 25 h, these two peaks show no relative decrease in intensity. It suggests that the short-range order of the BN nanosheets in the c-direction only slightly decreased after ball milling treatment, and the nanosheets were preferentially aligned with their in-plane surfaces parallel to the substrate. The latter effect should be due to the much increased aspect ratio of the nanosheets compared to that of the starting particles, resulting in the preferential "lying-down" position on sample substrates. This milling exfoliation method has also been successfully applied to produce other 2D nanomaterials, such as graphene and $MoS_2$ nanosheets.[65]

The room-temperature photoluminescence (PL) spectra of the BN nanosheets milled for different lengths of time are compared in Figure 8i.[60] The starting hBN powder (0 h) shows broad UV light emission centered at ~300 nm and a weaker DUV peak at 224 nm, typical of hBN materials. The BN nanosheets were milled in benzyl benzoate, bath sonicated, and annealed at 400 °C in vacuum, and a control sample was prepared by following a similar procedure without ball milling treatment. The PL results of the control sample show that these processes do not affect the light emission much. In contrast, BN nanosheets show greatly changed relative intensity between the DUV and UV peaks. The DUV light emission became relatively stronger after just 2 h of ball milling treatment and the relative intensity of the DUV peak increased with milling time up to 20 h. After 25 h of milling, new peaks located at 340 nm appeared.

The relative intensity change between the DUV and UV luminescence of the BN nanosheets should be due to their preferential orientation on the substrate. It is widely known that hBN has an anisotropic structure so that its luminescence is highly polarization anisotropic. The DUV light emission of hBN originates from the bandgap transitions between the valence and conduction bands with a $p_z$ character. Thus, the DUV luminescence is allowed only if the excitation light has a polarization perpendicular to the out-of-plane axis of the hBN crystals. The UV light emission is generally assigned to defects and disorders in hBN materials, which can have different polarization to that of the DUV emission. According to SEM investigations, the starting hBN particles were almost randomly oriented; however, after ball milling treatment, the nanosheets tended to lay flat on the substrate, as shown by the SEM image in Figure 8c and the XRD patterns in Figure 8h. Therefore, the change of the relative intensity between the DUV and UV light emissions is due to the preferential orientation of BN nanosheets, which have large diameter-to-thickness ratios.

The DUV light emission of the BN nanosheets also changed with ball milling time. The DUV luminescence of the BN nanosheets before and after the milling treatment is enlarged in Figure 8j.

The DUV light emission consists of three sub-peaks, at 217 nm, 225 nm, and 232 nm. These sub-peaks can be more clearly seen from the PL spectra at 30 K (Figure 8k). The 217 nm sub-peak corresponds to the lowest dipole-allowed Frenkel excitons with strong lattice interactions; the 225 nm and 232 nm sub-peaks can be assigned to excitons bound to stacking faults and structural defects.[55,66,67] The positions of the three sub-peaks show no noticeable shifts, indicating the absence of quantum confinement in the BN nanosheets. This could be due to the strongly bound and localized excitons in hBN so that quantum confinement is not present in BN nanosheets that are more than few layers thick.[67] The relative intensity of the three sub-peaks changes with ball milling time, as shown in Figure 8k. This can be ascribed to the increase of stacking faults in the BN nanosheets under the shear force during ball milling treatment. A later report confirmed that the luminescence of BN nanosheets has a weak dependence on thickness.[68]

## 7. Controlled Edge and Stacking in BN Nanoribbons

Nanoribbons can be regarded as a narrow strip form of nanosheets. Nanoribbons have many intriguing properties distinct from nanosheets, due to unique edge states and size effects. For example, the bandgap of graphene increases with the reduction of nanoribbon width.[69] Graphene nanoribbons also show novel optical, magnetic, and chemical properties.[70] The electronic structure of BN nanoribbons is much less width-dependent,[71] but edge effects can dramatically change the properties of BN nanoribbons. According to theoretical calculations, the bandgap of BN nanoribbons is tunable by electric field, and BN nanoribbons could be half-metallic with oxygen, hydrogen, fluorine edge terminations;[72] magnetism has also been predicted from BN nanoribbons.[71] In experiment, the electrical conductivity of BN nanoribbons was measured using TEM, and high conductivity was confirmed and explained by edge termination by oxygen atoms.[73] However, the synthesis of nanoribbons is challenging. Graphene nanoribbons have been produced via nanotube unzipping,[74-80] chemical routes,[81-84] and lithography.[85,86] BN nanoribbons were produced by unzipping BN nanotubes via plasma etching and alkali metal intercalation.[73,87] Nanotube unzipping is a popular method for the synthesis of both graphene and BN nanoribbons, in which two separate steps are required: nanotube synthesis and post-synthesis unzipping treatments.

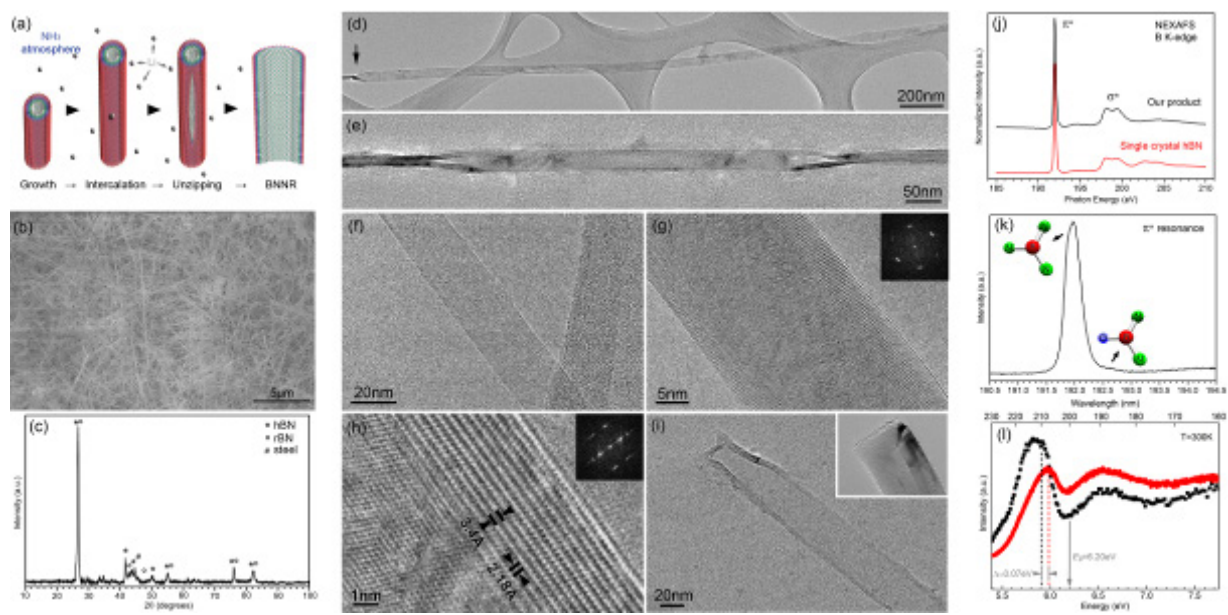

**Figure 9.** (a) The *in situ* unzipping process; (b) SEM image of product containing ~40% of BN nanoribbons; (c) XRD pattern of product; (d–i) TEM images of individual BN nanoribbons to show their unzipping site, thickness, stacking order, edge orientation, and tip; (j) NEXAFS spectra in B K-edge region of a single crystal of hBN and product; (k) enlarged NEXAFS spectrum of product; (l) comparison of PL excitation spectra of pure BN nanotubes and product containing 40% nanoribbons. Copyright © 2013 WILEY-VCH Verlag GmbH & Co. KGaA, Weinheim.[88]

The authors invented a one-step synthesis method of BN nanoribbons, i.e. unzipping of BN nanotubes during their synthesis (**Figure 9**a).[88] In the production process, amorphous B powder was ball milled and then mixed with $Li_2O$ powder and heated in ammonia gas at 1200 °C. The process is similar to the ball milling and annealing method for BN nanotube synthesis.[89-92] Figure 9b shows a SEM image of the product which contains many needle-like nanostructures. The XRD pattern in Figure 9c shows dominating hBN diffraction peaks, along with some rhombohedral BN phase and steel contaminations from the milling process. Although Li$_2$O was used, no Li phase was measurable by XRD or XPS.

According to TEM studies, the product contained about 40% of fully and partially unzipped BN nanoribbons (the rest were BN nanotubes). This yield is much higher than those of the previous methods.[73,87] The content of BN nanoribbons was further improved to about 60% using a non-optimized mild sonication and centrifugation method. Figure 9d shows a partially unzipped BN nanoribbon with the unzipping site arrowed on the left. Sometimes, two unzipping sites were observed on one partially unzipped BN nanoribbon, as illustrated in Figure 9e. It was found that the sonication and centrifugation process helped to fully unzip the nanoribbons, and these fully unzipped BN nanoribbons were down to 2L thin and more than 3 μm long. The width is normally in the range of 10–20 nm. TEM examination at higher magnifications revealed highly crystalline nanostructures. Figure 9g shows part of a 15L BN nanoribbon which has smooth edges. The inset shows the fast Fourier transform (FFT) pattern of the center of the nanoribbon (edges not included), indicating that it has the standard AA' stacking. Figure 9h shows the edge of the nanoribbon at higher magnification, and the 2.18 nm distance corresponds to (10−10) lattice spacing in hBN. It reflects that the BN nanoribbon has a well-ordered stacking. This is quite different from graphene nanoribbons, whose stacking is mostly random in the long range. The difference could be due to the additional ionic interaction between BN layers. The edge orientation of the BN nanoribbon was determined by the FFT pattern (inset in Figure 9h). Most of the BN nanoribbons had zigzag edges, and the rest are armchair oriented. In contrast, graphene nanoribbons usually do not have uniform edge orientation. The BN nanoribbons have narrowed ends (Figure 9i), because high-quality BN nanotubes normally have flat caps (inset of Figure 9i) where complete unzipping is difficult. Due to high flexibility, twists and folds were often observed in BN nanoribbons so that nanotubes and nanoribbons could sometimes be distinguished under SEM.

The quality and purity of the BN nanoribbons were analyzed using NEXAFS. The B K-edge NEXAFS spectra of an hBN single crystal and the BN nanoribbons are compared in Figure 9j. Both the single crystal and the BN nanoribbons show sharp π* resonances at 192.0 eV, representing the hexagonal B−3N bond. The width of the π* resonance for the two samples is similar, suggesting that the BN nanoribbons have a crystallinity comparable to the single crystal.[64] The enlarged π* resonance region of the BN nanoribbon shows only a very weak satellite peak at 192.6 eV, indicating a small amount of impurity in the form of a B−2N−O bond (Figure 9k).[62,63] There are no other impurities such as B−N−2O or B−3O in the nanoribbon sample. The high crystallinity and low level of impurity of the BN nanoribbons should be due to the high reactivity of Li, consistent with previous studies.[93,94] More importantly, the Li and NH$_3$ helped to *in situ* unzip the BN nanotubes to nanoribbons.[88]

The optical properties of the BN nanoribbons were also studied. Figure 9l shows the PL excitation spectra of pure BN nanotubes and a mixture of 40% nanoribbons and 60% nanotubes. The nanoribbons have a similar bandgap to nanotubes at 6.20 eV, which is slightly higher than that of BN nanosheets and hBN.[68] The excitation peak of the sample containing nanoribbons is located at 5.91 eV, which is 0.07 eV less than that of the nanotubes. In other words, BN nanoribbons have a larger exciton binding energy than nanotubes. Some properties and applications of graphene and BN nanosheets are compared in **Table 1**.

Table 1. Comparison of properties and applications of graphene and BN nanosheets

| Property/application | Graphene | BN nanosheets |
|---|---|---|
| Bond length | 1.42 Å | 1.44 Å |
| Bandgap | None | ~6 eV[4] |
| Young's modulus | 1.0 TPa[9-10] | 0.71–0.97 TPa (theoretical)[5-8] |
| Breaking strength | 130 GPa[9-10] | 120–165 GPa (theoretical)[5-8] |
| Thermal conductivity | 1800–5400 W m$^{-1}$ K$^{-1}$ [95,96] | 100–270 W m$^{-1}$ K$^{-1}$ (few-layer)[11-13] |
| Oxygen doping | 250 °C[17] | >700 °C[22] |
| Etching by oxidation | 450 °C[17] | >800 °C[22] |
| Galvanic corrosion | Yes[31-32] | No[33] |
| Protection barrier | No[31-32] | Yes[33] |
| Reusable surface-enhanced Raman spectroscopy substrate | No | Yes[19] |
| Dielectric screening | Highly thickness dependent[46] | Less thickness dependent[45] |
| Luminescence | No | Up to DUV region[56] |

## 8. Conclusions and Outlook

The results described here highlight the special properties and applications of BN nanosheets. In contrast to the oxidation of graphene at 400 °C, high-quality 1L BN can resist oxidation at temperatures above 800 °C. The oxidation temperature only slightly decreases with the increasing level of defects in BN nanosheets. BN nanosheets are thus more suitable for high-temperature applications. High thermal stability and impermeability make BN nanosheets potential candidates for metal protection. More importantly, where graphene promotes the corrosion of underlying metal due to the formation of a galvanic cell, BN nanosheets are electrically-insulating and hence do not cause galvanic corrosion. Covered with plasmonic Au nanoparticles, BN nanosheets are effective and reusable substrates for SERS to detect chemicals at low concentrations. BN nanosheets can also serve as excellent dielectric substrates for graphene and other 2D nanomaterials to form heterostructures for electronic and optical applications. The dielectric screening in BN nanosheets has a relatively weak dependence on thickness, which is very different to that of graphene. BN nanosheets are better DUV light emitters compared to hBN particles owing to the preferential orientation of the nanosheets on substrate, which is preferable for polarization anisotropic light emission. BN nanoribbons produced by the *in situ* unzipping method have a uniform stacking order and mostly zigzag edges. Due to the edge effect, BN nanoribbons show a larger exciton binding energy than nanotubes, in spite of the similar bandgap of the two BN nanostructures.

There is no doubt that more special properties and applications will be revealed and realized from monolayer and few-layer BN nanosheets, but the most interesting discoveries must be those exclusive to BN nanosheets and unavailable to graphene and other 2D nanomaterials. Many unique and exciting properties have been theoretically predicted for 1L BN and await experimental verification, so they can act as guidance for future research in the field. High-quality BN nanosheets are prerequisite for these studies because they not only facilitate the exploration of their intrinsic properties but also are critical for many applications, including the oxidation and corrosion protection mentioned in this article. Nevertheless, the synthesis of large BN nanosheets with low defects is currently much less sophisticated than that of graphene. Therefore, it is of great importance to improve the synthesis of BN nanosheets of high quality and large size via both top-down exfoliation[97] and bottom-up fabrication by CVD and chemical reaction methods.[98] Another promising research direction is the modification of BN nanosheets and nanoribbons via doping and functionalization, or production of hybridized nanosheets containing BN phases. The difficulty lies in precise experimental control over the doping site, concentration, and fine structure in the hybrid.

## 9. Experimental Section

BN nanosheets of relatively low quality were mechanically exfoliated from commercial hBN particles (PT110, Momentive). To test their resistance to oxidation, the nanosheets were heated in a horizontal tube furnace at different temperatures in air for 2 h. The AFM images were acquired using silicon cantilevers (spring constant 40 N m$^{-1}$) in tapping mode.

For the metal protection test, 1L CVD-grown BN nanosheets on Cu foil were purchased (Graphene Supermarket) and used as received. The procedures for the heating treatments at 250 °C in air and electrochemical tests in 0.1 M NaCl solution (0.1 M) were exactly the same as has been previously described in the literature[33]. The characterization details also refer to this reference.

**Acknowledgements**

L.H. Li thanks the Alfred Deakin Postdoctoral Research Fellowship grant for financial support. Financial support from the Australian Research Council under the Discovery Program and DECRA (DE160100796) is gratefully acknowledged. Part of this research was undertaken on the soft X-ray beamline at the Australian Synchrotron, Victoria, Australia.

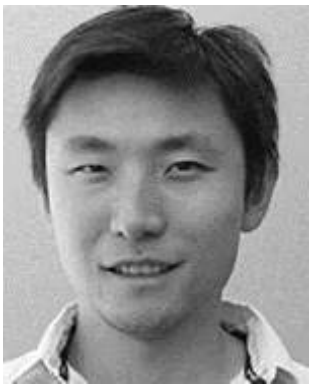

**Lu Hua Li** received his B.Sc. degree from Beijing Institute of Technology, China (2003) and Ph.D. degree from the Australian National University, Australia (2012). He is currently working at Deakin University, Australia under ARC Discovery Early Career Researcher Award (DECRA). His research interest lies in synthesis, advanced characterization, properties, and applications of 2D nanomaterials, especially BN nanosheets.

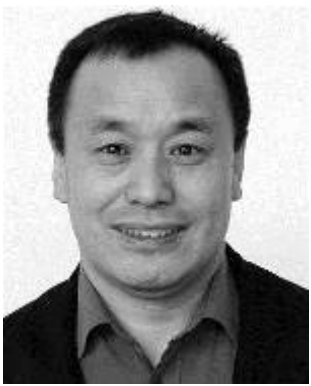

**Ying Chen** is an Alfred Deakin professor and the chair in nanotechnology at Deakin University, Australia. He obtained his PhD degree from University Paris Sud. His current research focuses on the development of different synthetic methods (ball milling and annealing method, chemical/physical vapour deposition, and mechanical alloying) and new applications (energy storage in batteries, capacitors, solar cells, environmental protection, and drug delivery) of a variety of nanomaterials, including boron nitride nanotubes and nanosheets, graphene, and nanowires, nanorods, nanoparticles, and thin films of oxides and nitrides.